\documentclass[a4paper,pre,nofootinbib,twocolumn,showpacs]{revtex4}
\usepackage[dvips]{epsfig}
\usepackage[dvips]{graphicx}
\usepackage{latexsym,amsmath,verbatim,amssymb}
\usepackage[usenames,dvipsnames]{color}
\newcommand{\be}{\begin{equation}}
\newcommand{\ee}{\end{equation}}

\usepackage{array}

\begin{document} 
\title{Quenched disorder forbids discontinuous transitions in
  non-equilibrium low-dimensional systems}

\author{Paula Villa Mart{\'\i}n}

\affiliation{Departamento de Electromagnetismo y F\'\i sica de la
  Materia, Facultad de Ciencias, Universidad de Granada, 18071
  Granada, Spain }

\author{Juan A. Bonachela} \affiliation{Department of Ecology and
  Evolutionary Biology, Princeton University, Princeton, NJ,
  08544-1003, USA}
 
\author{Miguel A. Mu\~noz}
\affiliation{Departamento de Electromagnetismo y F\'\i sica de la
  Materia and Instituto Carlos I de F\'\i sica Te\'orica y
  Computacional, Facultad de Ciencias, Universidad de Granada, 18071
  Granada, Spain }

\date{\today}

\begin{abstract}
  Quenched disorder affects significantly the behavior of phase
  transitions.  The Imry-Ma-Aizenman-Wehr-Berker argument prohibits
  first-order/discontinuous transitions and their concomitant phase
  coexistence in low-dimensional equilibrium systems in the presence
  of random fields. Instead, discontinuous transitions become rounded
  or even continuous once disorder is introduced.  Here we show that
  phase coexistence and first-order phase transitions are also
  precluded in non-equilibrium low-dimensional systems with quenched
  disorder: discontinuous transitions in two-dimensional systems with
  absorbing states become continuous in the presence of quenched
  disorder.  We also study the universal features of this
  disorder-induced criticality and find them to be compatible with the
  universality class of the directed percolation with quenched
  disorder. Thus, we conclude that first order transitions do not
  exist in low-dimensional disordered systems, not even in genuinely
  non-equilibrium systems with absorbing states.
\end{abstract}

\pacs{ 05.70.Ln, 02.50.Ey, 64.60.Ht  }

\maketitle
\section{Introduction}

Quenched disorder has a dramatic effect on both the statics and the
dynamics of phase transitions
\cite{Harris-Lubensky,GG-Luther,Imry-Wortis}.  A time-honored argument
by Imry and Ma explains in a simple and parsimonious way why
symmetries cannot be spontaneously broken in low-dimensional systems
in the presence of quenched random fields \cite{Imry-Ma}.  In a
nutshell, the argument is as follows.  Suppose a discrete symmetry
(e.g., $Z_2$ or up-down) was actually spontaneously broken in a
$d$-dimensional system and imagine a region of linear size $L$ with a
majority of random fields opposing the broken-symmetry state. As a
direct consequence of the central limit theorem, by reversing the
state of such a region the bulk-free energy would decrease
proportionally to $L^{d/2}$, but this inversion would also lead to an
interfacial energy cost proportional to $L^{d-1}$. Comparing these two
opposing contributions for large region sizes, it follows that for $d
\leq 2$ the first dominates, making the broken-symmetry state
unstable. If the distinct phases are related by a continuous symmetry,
soft modes reduce the effect of the boundary conditions to $L^{d-2}$
and the marginal dimension is $d =4$ \cite{Aizenman}. Thus, the
energetics of low-dimensional systems is controlled by the random
field, which is symmetric, thus preventing symmetries from being
spontaneously broken and continuous phase transitions from
existing. Instead, in higher dimensional systems, the situation is
reversed, symmetries can be spontaneously broken, and phase
transitions exist.

The Imry-Ma argument {\it i)} holds for equilibrium systems (where the
free energy is well defined), {\it ii)} is backed by more rigorous
renormalization group calculations, which prove that no
symmetry breaking occurs even at the marginal case $d=2$ (where rough
interfaces could potentially break the argument above
\cite{Aizenman}),  {\it iii)} has been verified in countless
examples both experimentally and numerically, and (iv) has been
extended to quantum phase transitions \cite{Joel,Vojta-Quantum}.

In contrast with the equilibrium case, recent work by Barghathi and
Vojta \cite{Vojta12}, shows that second-order phase transitions may
survive to the introduction of random fields even in one-dimensional
cases \cite{Masuda10,Pigolotti-Cencini} in genuine non-equilibrium
systems with absorbing states for which there is not such a thing as
free energy
\cite{Marro-Dickman,Haye_Book,Odor-Book,GG-Granada}. Therefore, {\it
  the Imry-Ma argument does not apply to these non-equilibrium systems
  owing to the presence of absorbing states, and, in consequence,
  states of broken symmetry can exist in the presence of random
  fields.}

Let us now shift the discussion to first-order phase transitions, for
which system properties such as the magnetization, energy, density,
etc., change abruptly as a control parameter crosses a threshold value
at which two distinct phases coexist.  As shown by Aizenman and Wehr,
first-order phase transitions in low-dimensional equilibrium systems
are rounded (made less sharp) by disorder, and, even more remarkably,
the rounding may result into a critical point; i.e.,
first-order/discontinuous phase transitions become
second-order/continuous ones upon introducing (random-field) disorder
\cite{Aizenman}.  A similar conclusion applies to the case of random
interactions \cite{Aizenman,Berker,Hui-Berker}; indeed, a random
distribution of interactions (e.g., bonds) locally favors one of the
two phases, and thus, it has the same effects as random fields.
Different Monte Carlo results support this conclusion; furthermore
they suggest that the disorder-induced continuous transition exhibits
critical exponents which are consistent with those of the
corresponding pure model. An argument explaining these findings was
put forward by Kardar {\it et al.} \cite{Kardar}.

In close analogy with the argument above for the absence of
symmetry-breaking, in the case of phase coexistence as well,
regions (or ``islands'') of arbitrary size of one of the phases appear
in a stable way within the other.  Therefore, islands exist within
islands in any of the two phases in a nested way, leading always to
hybrid states. Hence, two distinct phases cannot possibly coexist and
first-order transitions are precluded in disordered low-dimensional
equilibrium systems.

Thus, the question arises as to whether shifting to the
non-equilibrium realm entails the shattering of a fundamental
cornerstone of equilibrium statistical mechanics as it happens for
continuous phase transitions (see table 1 for a synthetic summary);
do first-order phase transitions, and, hence, phase coexistence,
exist in low-dimensional non-equilibrium disordered systems?

Aimed at shedding some light on this issue, we study non-equilibrium
models with absorbing states in the presence of disorder. More
specifically, we study a variant of the well-known ``contact process'',
sometimes called the ``quadratic contact process'', in which two
particles are needed to generate an offspring while isolated particles
can spontaneously disappear
\cite{Liggett,Marro-Dickman,Haye_Book,Odor-Book,Durret_QCPnetworks}. As
a first step, we verify that the pure version of the model exhibits a
first-order transition separating an active phase from an absorbing
one. Then we introduce disorder in the form of a site-dependent
transition rates and investigate whether the discontinuous character
of the transition survives.

\setlength{\tabcolsep}{2.3pt}
\renewcommand{\arraystretch}{3}
\newcolumntype{C}[1]{>{\centering\let\newline\\\arraybackslash\hspace{0pt}}m{#1}}

\begin{center}
\begin{table} [h!]
\begin{tabular}{ | C{3cm} || C{2.5cm} | C{2.5cm} |}
    \cline{2-3}
   \multicolumn{1}{C{3cm}||}{\textbf{System with Random Fields
       \newline $d\le 2$  }}
    & \textbf{{\color{blue}$2^{nd}$ order} \newline (\small{spontaneous sym. breaking})}
    & \textbf{{\color{blue}$1^{st}$ order} \newline (\small{phase coexistence})} \\
    \hline
    \hline
    {\textbf{{\color{blue} Equilibrium}}} & {\color{Gray}NO \cite{Imry-Ma}} & {\color{Gray}NO \cite{Imry-Ma,Aizenman,Berker,Hui-Berker}} \\ 
    \hline
    {\textbf{{\color{blue} Non-equilibrium }\newline (abs. states)}} & {\color{Gray}YES \cite{Vojta12}} & {\color{Red}\textbf{\Huge{?}}} \\ 
    \hline
\end{tabular}
\caption{\textbf{Random fields in low-dimensional disordered systems.} 
  Summary of the effects of quenched random  fields on the existence
  of continuous/second-order transitions (with spontaneously symmetry breaking), 
  and discontinuous/first-order (with associated phase coexistence)
  phase transitions in $d\le2$ systems. Both, the equilibrium and
  non-equilibrium cases are considered, the latter including the possibility of one or more absorbing states. }
\end{table}
\end{center}

\section{Two possible scenarios}

Two alternative scenarios might be expected {\it a priori} for the
impure/disordered model:
\begin{enumerate}
\item the Imry-Ma argument breaks down in this non-equilibrium case
  and a {\it first order  phase transition} is observed, or
\item the Imry-Ma prediction holds even if the system is a
  non-equilibrium one, and a disorder-induced {\it second-order phase
    transition} emerges.
\end{enumerate}

If the latter were true, we could then ask what universality class
such a continuous transition belongs to. A priori, it could share
universality class with other already-known critical phase transitions
in disordered systems with absorbing states
\cite{Hoo1,Cafiero,Vojta1D,Vojta2D,Vojta-Review} or, instead, belong
to a new universality class defined by this disorder-induced
criticality.

If no novel universal behavior emerges, then it is expected for the
model to behave as a standard two-dimensional contact process (or
directed percolation) with quenched disorder with the following main
features \cite{Hoo1,Vojta1D,Vojta2D,Vojta-Review}:
\begin{itemize}

\item
there should be a
critical point separating the active from the absorbing phase,

\item at criticality, a
logarithmic or {\it activated} type of scaling (rather than algebraic)
should be observed. For instance, for quantities related to activity
spreading such as the survival probability, averaged number of
particles, and radius from a localized initial seed, we expect $
P_s(t) \sim [\ln(t/t_0)]^{-\bar{\delta}}$, $N(t) \sim
[\ln(t/t_0)]^{\bar{\theta}}$, and $R(t) \sim [\ln(t/t_0)]^{1/{\Psi}}$,
respectively; $t_0$ is some crossover time, and $\bar{\delta}$,
$\bar{\theta}$, and $\Psi$ should take the values already reported in
the literature \cite{Vojta2D}.

\item there should be a sub-region of the absorbing phase, right below
  the critical point, exhibiting generic algebraic scaling with
  continuously varying exponents, i.e. a Griffiths phase
  \cite{Griffiths}. Griffiths phases stem from the existence of
  rare regions where the disorder takes values significantly different
  from its average \cite{Vojta-Review}.
\end{itemize}

These features follow from a strong-disorder renormalization group
approach for the disordered contact process, which concludes that this
anomalous critical behavior can be related to the random
transverse-field Ising model for sufficiently strong disorder
\cite{Hoo1}, and have been confirmed in computational studies which
suggested that this behavior is universal regardless of disorder
strength \cite{Vojta2D,Vojta-Review,Hoyos2008}.

\section{Model and Results} 

We study the simplest non-equilibrium model with absorbing states
exhibiting a first-order/discontinuous transition. Given that, owing
to different reasons, one-dimensional systems with absorbing states
rarely exhibit first-order phase transitions (even in pure systems)
\cite{Haye1,Haye2}, here we focus on the physically more relevant
two-dimensional case. In this two-dimensional reaction-diffusion
contact-process-like model \cite{Marro-Dickman,Haye_Book,Odor-Book,GG-Granada}, individual
particles disappear at a fixed rate, $\mu$, while a pair of
nearest-neighbor particles is required to create an offspring at some
rate $\lambda$
\begin{equation}\label{model}
  A \stackrel{\mu}{\rightarrow} \emptyset  \,,\quad \,\,
  2 A \stackrel{\lambda}{\rightarrow} 3 A  \,,\quad \,
%\,
%  3 A \stackrel{\sigma}{\rightarrow} 2 A  \,.
\end{equation}
with the additional (``hard-core'' or ``Fermionic'') constraint preventing sites 
from housing more than one particle. This restriction can be relaxed
at the cost of introducing a reaction
\begin{equation}\label{model2}
  3 A \stackrel{\lambda}{\rightarrow} 2 A  \,,
\end{equation}
which keeps the number of particles bounded.  In either case, the
corresponding rate or mean-field equations are
(see Appendix)
\begin{equation}
\dot{\rho(t)} = - \mu \rho(t) + \lambda \rho(t)^2 (1- \rho(t)),
\label{rate}
\end{equation}
where $\rho$ represents the density of active sites or particles. This equation has the trivial stationary solution, $\rho=0$ and an additional
one at $\rho^*=\frac{1}{2} ( 1 + \sqrt{1 -4 \mu/\lambda})$ for $\lambda>4 \mu$, with an associated discontinuous transition at
$\lambda=4 \mu$.

\subsection{Pure model}

Among the many possible ways in which the above particle system can be
implemented \cite{Oliveira_model,Evans_model2,Jensen_model}, we employ
the model proposed in Ref. \cite{Jensen_model}, which was 
numerically studied  in two-dimensions and verified to exhibit a first-order phase
transition separating an active from an absorbing phase
\cite{Jensen_model}.

We consider a two-dimensional square lattice and define a binary
occupation variable $s=0,1$ (empty/occupied) at each site.  We
consider some initial conditions and perform a sequential updating
following the standard procedure
\cite{Marro-Dickman,Haye_Book,Odor-Book,GG-Granada}: {\it i)} an
active site is randomly selected (from a list including all active
ones); {\it ii)} with probability $p_d$ (death) the particle is
annihilated, otherwise, with complementary probability $1-p_d$ a
nearest neighbor site is chosen; {\it iii)} if this latter is empty,
the selected particle diffuses to it, and otherwise an offspring particle
is created at a randomly chosen neighboring site with probability
$p_b$ (birth) provided it was empty; otherwise nothing happens. We
keep $p_b=0.5$ fixed and use $p_d$ as the control parameter.

As customarily done, we perform two types of experiments
\cite{Marro-Dickman,Haye_Book,Odor-Book,GG-Granada}, considering as
initial condition either a homogeneous state, i.e., a fully occupied
lattice of linear size $L$, or a localized seed, consisting in this
case of a few, at least a couple, neighboring particles in an
otherwise empty lattice.

\begin{figure}
%\begin{center}
  \includegraphics[width=0.485\textwidth]{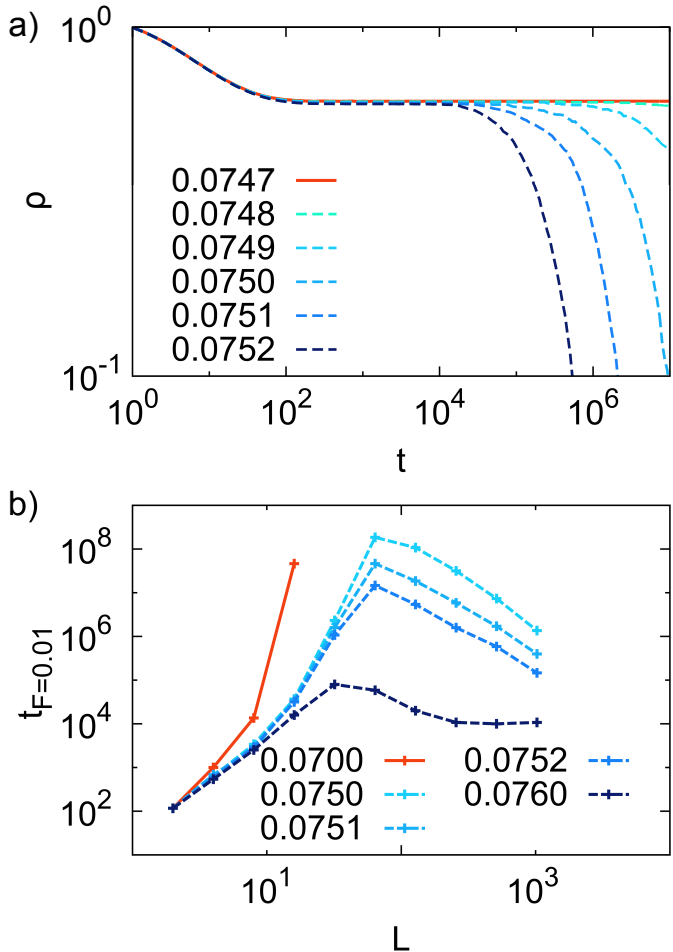}
  \caption{\textbf{(Color online) Decay from a homogeneous initial
      condition in the pure model }. (a) Time evolution of the total
    averaged particle density for $N=L^2=256^2$; a first order
    phase transition can be observed near $p_{d_{thr}} \approx
    0.0747$.  (b) Mean survival time, $t_F$, required to reach an
    arbitrarily small density value here fixed to $0.01$ (results are
    robust against variations of this choice) as a function of system
    size. Up to $10^3$ realizations have been used to average these
    results. From this finite size analysis, the threshold point can
    be bounded to lie in the interval $[0.070, 0.075]$.}
\label{fig:1}
%\end{center}
\end{figure}

{\it Homogeneous initial conditions--} Figure~\ref{fig:1} shows
results of computer simulations for the temporal decay of the particle
density from $\rho(t=0)=1$. The upper panel shows an abrupt change of
behavior at a threshold value $p_{d_{thr}} \approx 0.0747$; activity
survives indefinitely for $p_d< p_{d_{thr}}$ (at least up to the
considered maximum time) and the particle density converges to
relatively large steady state values ($\rho \approx 0.6$), while
activity dies off exponentially for any $p_d>p_{d_{thr}}$.  This
behavior is compatible with a first order phase transition, but 
the location of the threshold value has to be considered as a rough
estimate.

To better locate the transition point, we study the mean survival time
(MST) as a function of system size.  Figure~\ref{fig:1}b
shows a non-standard non-monotonous dependence of the MST as a
function of size $N= L^2$.  As we see, there are two regimes: {\it i)} for
$L<L_c$ there is an exponential increase of the MST with system size;
{\it ii)} for $L_c<L$, and quite counter-intuitively, the MST
decreases with increasing system size. This behavior can be
rationalized following recent work where a particle system very
similar to ours is studied by employing a semiclassical approach
\cite{Meerson} (see also Ref. \cite{Lubensky}). Following this study, the
first regime corresponds to the standard Arrhenius law, i.e., the fact
that a quasi-stationary state with a finite particle density
experiences a large fluctuation extinguishing the activity in a
characteristic time which grows exponentially with system size
\cite{Gardiner}. On the other hand, there is a ``critical system
size'' above which the most likely route to ``extinction'' consists on
the formation of a critical nucleus that then expands
in a ballistic way, destabilizing the quasi-stationary state. 
Obviously, the larger the system size the most likely that a critical
nucleus is spontaneously formed by fluctuations. Finally, for
sufficiently large system sizes there is a last ``multi-droplet''
regime in which many nuclei are formed and the MST ceases to depend on
system size, reaching and asymptotic value \cite{Meerson}.  This
picture fits perfectly well with our numerical findings.

From this analysis, we conclude that, with the present computational
resolution, we can just give a rough estimation for the location of
the transition point $ 0.070 < p_{d_{thr}} < 0.075$.

\begin{figure}
  \centering
  \includegraphics[width=0.48\textwidth]{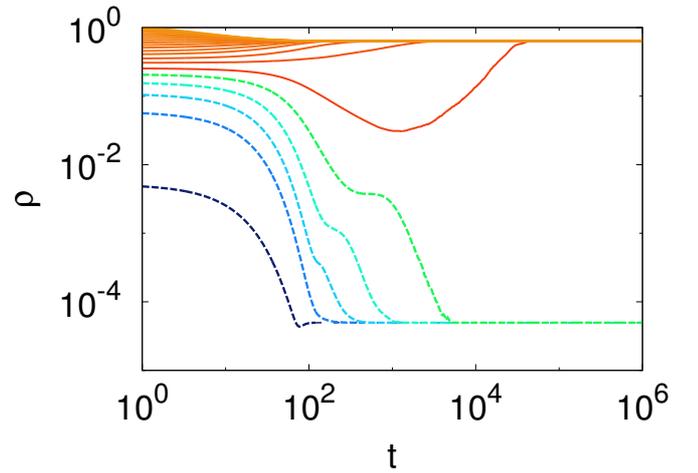}
  \caption{{\textbf{(Color online) Bistability at the transition point
        of the pure model}.  Log-log plot of the averaged particle
      density as a function of time for different initial conditions
      in the neighborhood of the transition point (results here are
      for $p_d=0.07315$). Depending on the initial density, the system
      stabilizes in the active or in the absorbing phase. The selected
      initial densities are equi-spaced in the interval $[0.005,1]$
      with constant increments $0.05$; system size $N=256^2$ and
      averages performed over up to $10^6$ realizations.}}
  \label{fig:3}
\end{figure}

To show further evidence of the discontinuous nature of the phase
transition, figure~\ref{fig:3} illustrates the system bistability
around the transition point: depending on the density of the initial
configuration, a homogeneous steady state may converge either
to a stationary state of large density (active) or to the absorbing
state. A separatrix marks the distinction between the two different
basins of attraction.  Let us remark that systems exhibiting a first-order
transition are bistable only at exactly the transition point
but for finite system sizes the coexistence region has some
non-vanishing thickness.  The existence of bistability makes a strong
case for the discontinuous character of the transition.

\begin{figure}
  \centering
  \includegraphics[width=0.48\textwidth]{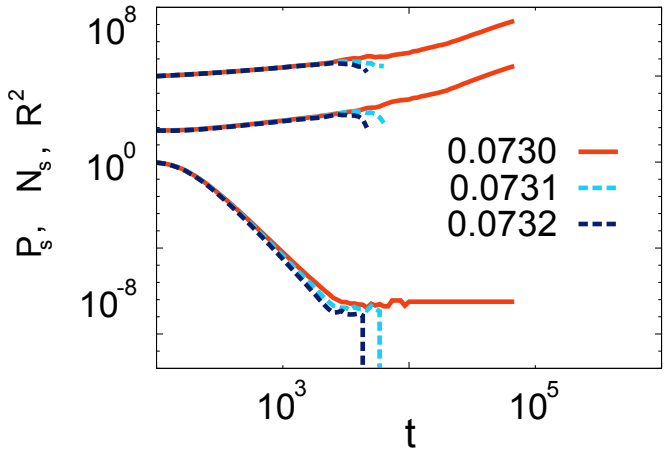} 
  \caption{\textbf{(Color online) Spreading experiments for the pure
      model}.  Double-logarithmic plot of the (from bottom to top),
    (i) the survival probability $P_s(t)$, (ii) the averaged number of
    particles $N_s(t) $ (averaged over surviving runs), and (iii) the
    averaged squared radius (averaged over of all runs), as a function
    of time, for $N=1024^2$ using up to $4\times 10^{10}$ experiments.
    Curves for $R^2(t) $ have been shifted upwards for clarity.  In
    spite of the large number of runs used to average, curves are
    still noisy.  This is due to the fact that, being very close to
    the transition point, a large fluctuation is needed for the system
    to "jump" to the active phase from the vicinity of the absorbing
    one and only a few runs reach large times.}
  \label{fig:2}
\end{figure}

{\it Spreading experiments from a localized seed--} We consider a few
(at least $2$) neighboring particles at the center of an otherwise
empty lattice, and monitor how activity spreads from that seed. Each
simulation run ends whenever the absorbing state is reached or when
activity first touches the boundary of the system. We monitor the
averaged squared radius from the origin $R^2(t)$, the averaged number
of particles over surviving trials, $N_s(t)$, and the survival
probability, $P_s(t)$ \cite{Marro-Dickman}. Figure~\ref{fig:2} shows
log-log plots of these three quantities as a function of time. In all
cases, we find a threshold value $p_{d_{thr}} \approx 0.073$ that
marks a change of tendency, signaling the frontier between the
absorbing and active phases. In the active phase ($p_d<p_{d_{thr}}$)
and for large values of $t$, both $N_s(t)$ and $R^2$ grow
approximately as $t^2$ (as expected for ballistic expansion), while $P_s(t)$ converges to a constant (i.e. some runs do
survive indefinitely).  On the other hand, in the absorbing phase all
three quantities curve downwards indicating exponential extinction.

Thus, the pure model exhibits a discontinuous transition at some value
of $p_{d_{thr}} \approx 0.073$, which separates a phase of high activity from an
absorbing one. Observe, that the estimation of the transition point is
compatible with the interval obtained above. 
  
\subsection{Disordered model}

In the disordered version of the model, each lattice site has a random
uncorrelated (death) probability. In particular, we take $p_d({\bf
  x})= p_d r $ where $p_d$ is a constant and $r$ is a
homogeneously-distributed random number $ r\in [0,2]$ (and, thus, the
mean value is $p_d$). Spatial disorder is refreshed for each run, to
ensure that averages are independent of any specific realization of
the disorder.

\begin{figure}
         \begin{center}
         \includegraphics[width=0.47\textwidth]{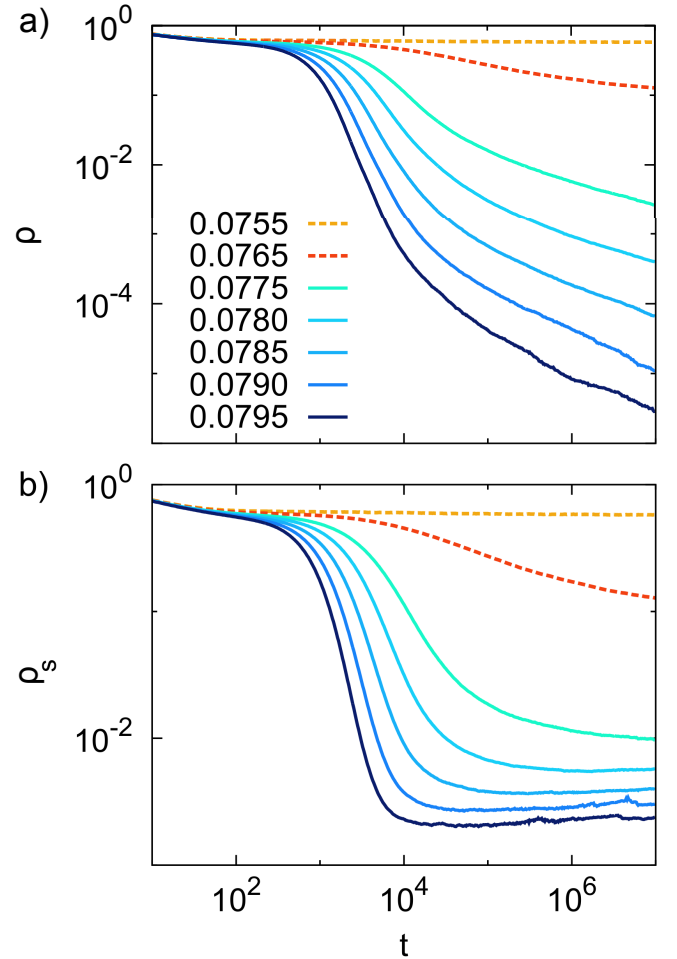}
         \caption{\textbf{(Color online) Density decay from a
             homogeneous initial conditions in the disordered model.}
           (a) Particle density averaged over all trials in a lattice
           of size $256^2$ and up to $2\times 10^4$ realizations
           (curves in the active phase are plotted with dashed
           lines). Observe the presence of a broad region with generic
           power-law behavior, i.e. a Griffiths phase which starts
           roughly at $p_d=0.0775$. (b) As (a) but averaging only over
           surviving trials. Note the non-monotonic behavior in the
           Griffiths phase (see main text for details). }
\label{fig:4}
\end{center}
\end{figure}

{\it Homogeneous initial conditions--} We have computed time series
for {\it i)} the mean particle density averaged over all runs and {\it
  ii)} the mean particle density for surviving runs (i.e. those which
have not reached the absorbing state). Figure~\ref{fig:4} shows time
evolution after up to $2\times 10^{4}$ realizations. Results are
strikingly different from those of the pure-model.  

For values below threshold, $p_d<p_{d_c}\approx 0.077$, the particle
density converges to a constant value for asymptotically large times,
while for $p_{d_c} > 0.077$ curves decay as power laws (a much more
precise estimation of the critical point will be computed below).  The
generic algebraic decay is observed for a wide range of $p_d$;
however, the transient before the power-law regime increases with
$p_d$, which makes it difficult to determine the exact boundaries of
the mentioned range.  The presence of generic algebraic scaling in an
extended region is the trademark of Griffiths phases.

Plotting the activity over the surviving trials [Fig. ~\ref{fig:4}b],
we observe that the evolution is non-monotonous in the absorbing
(Griffiths) phase: the curves decrease up to a minimum value and then
increase.  This stems from the fact that realizations with large rare
active regions remain active for longer times than those with smaller
ones; as realizations with only relatively small rare-regions
progressively die out, those with larger and larger rare-regions are
filtered through and, thus, the overall average density grows as a
function of time, being limited only by system size. 

\begin{figure}[h]
         \centering
    \includegraphics[width=0.48\textwidth]{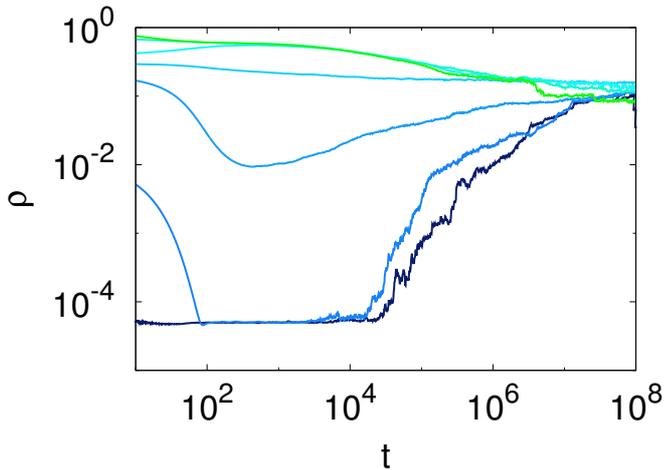}
    \caption{\textbf{(Color online) Absence of bistability in the
        disordered model.} Double-logarithmic plot of the averaged
      particle density as a function of time for $p_d=0.0765$, with
      $N=256^2$, and up to $10^6$ realizations.  Initial densities are
      $\rho_0={0.00006,0.01,0.2,0.3,0.4,0.7,1}$. Regardless of the
      initial condition, the system stabilizes to a constant small
      value of the density, as expected for a second-order phase
      transition.}
\label{fig:5} 
\end{figure}

In addition, we observe that, contrarily to the pure case, there is no
bistability around the transition point (figure~\ref{fig:5}). Indeed,
very near to the transition point ($p_d=0.07650$), all curves
regardless of their initial value converge to a unique well-defined
stationary density close to zero, as appropriate for a continuous
transition to an absorbing state.

\begin{figure}[h!] 
         \centering
         \includegraphics[width=0.48\textwidth]{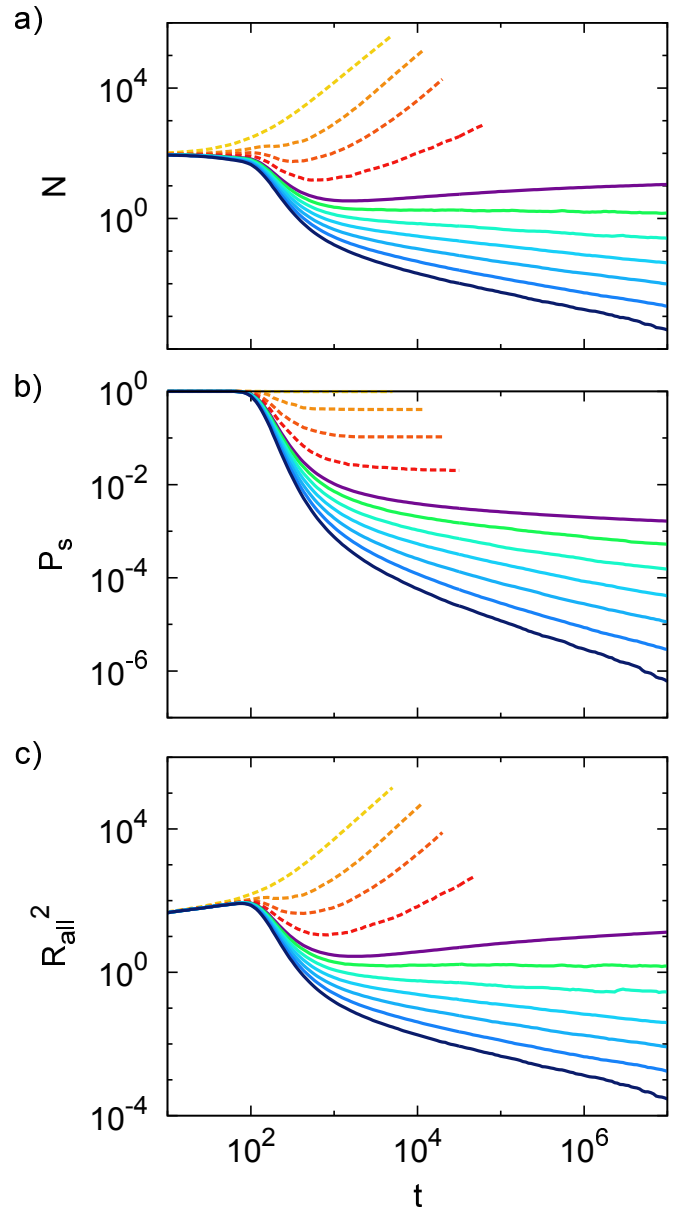}
         \caption{\textbf{(Color online) Spreading experiments in the
             disordered model}. Double logarithmic plot of the three
           usual spreading quantities showing the presence of generic
           power-laws with continuously varying exponents all along
           the Griffiths phase ($p_d \lesssim 0.07652$).  Parameter
           values (from top to bottom) $0.06,0.07, 0.073, 0.075,
           0.07652,0.077, 0.0775,$ $0.078, 0.0785, 0.079, 0.0795$
           (curves in the active phase are plotted with dashed lines),
           up to $5\times 10^7$ realizations.}
           \label{fig:6}
\end{figure}

{\it Spreading experiments from a localized seed--} Figure~\ref{fig:6}
shows results for three spreading observables as a function of time;
for all of them, we clearly observe generic asymptotic power laws with
continuously varying exponents.
These spreading quantities also allow us to scrutinize the behavior at
the critical point. As discussed in the Introduction, in a disordered
system as the one under study, we expect logarithmic (activated
scaling) at criticality. Indeed, Figure 7 shows results for the usual
spreading quantities represented in a double logarithmic plot of the
different quantities as a function of $ln(t/t_0)$. The value of $t_0$
is in principle unknown and constitutes a significant error source
\cite{Vojta2D}. We fix it as the value of $t$ such that it gives the
best straight lines at the transition point for all three quantities
\cite{Vojta2D}).  Right at the critical point ($p_c \approx 0.07652$
to be obtained with more accuracy below) a straight asymptotic
behavior indicates that results are compatible with logarithmic
(i.e. activated) type of scaling.  The best estimates for the
(pseudo)-exponents listed in section II are: $\bar{\delta} \approx
{1.90}$, $\bar{\theta} \approx {2.09}$, and $\Psi \approx {0.43}$,
which are compatible with the values reported in the literature
for the universality class of directed percolation with quenched
disorder (i.e.  $\bar{\delta} =1.9(2)$, $\bar{\theta} = 2.05(20)$,
$\Psi= 0.51(6)$).

Similarly, following the work of Vojta and collaborators
\cite{Vojta2D}, we represent in Fig. 8 one of the spreading
quantities as a function of another one, e.g. $N(t)$ as a function of
$P_s(t)$ to eliminate the free variable $t_0$ from the plot.  This
type of plot allows for the identification of power-law dependencies
rather than logarithmic ones, i.e. $N(t) \sim
P_s(t)^{-\bar{\theta}/\bar{\delta}}$. If the second-order phase
transition belongs to the universality class of the directed
percolation with quenched disorder (see section II), we should have
$N(t) \sim P_s(t)^{-1.08(15)}$, using as a reference the values in the
literature \cite{Vojta2D}. Indeed, as shown in Figure 8 we obtain
$N(t) \sim P_s(t)^{-1.10(2)}$, in very good agreement with the
expected value \cite{Vojta2D}, and this is the method by which the
critical point location, $p_d \approx 0.07652$, is obtained with best
accuracy.

\begin{figure}[h!]
         \centering
         \includegraphics[width=0.48\textwidth]{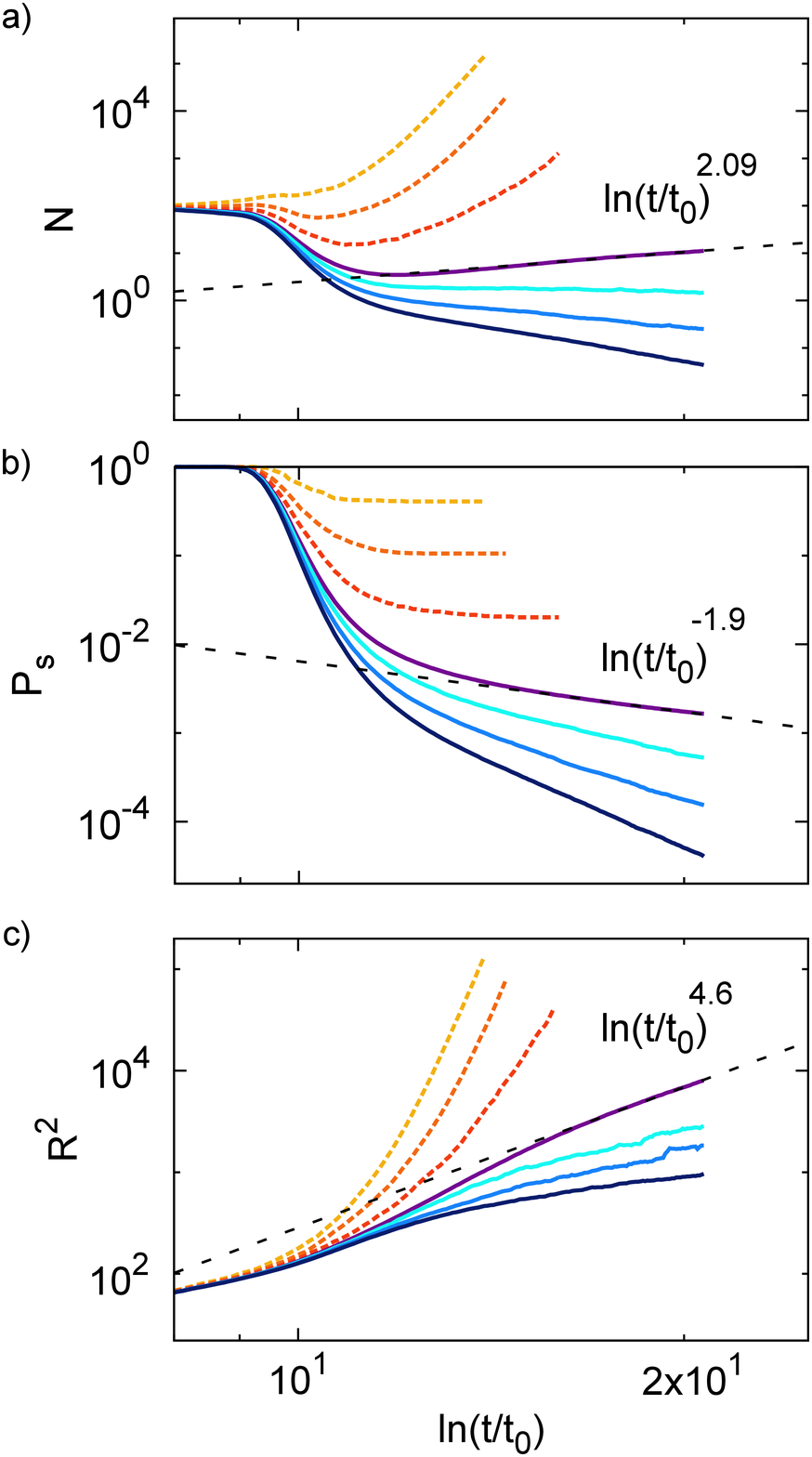}
         \caption{\textbf{(Color online) Spreading experiments in the
             disordered model}. Double logarithmic plot of the three
           usual spreading quantities as a function of $\ln(t/t_0)$
           for different parameter values (from top to bottom $0.07,
           0.073, 0.075, 0.07652, 0.077,$ $0.0775, 0.078$ (curves in
           the active phase are plotted with dashed lines).  Same
           network sizes as in the homogeneous case and up to $10^6$
           realizations. By conveniently choosing $t_0 =0.01$ (see
           main text) we observe straight lines at the critical point,
           $p_d \approx 0.07652 $.  From their corresponding slopes we
           measure the associated (pseudo)-exponents: $\bar{\theta}
           \approx 2.09$, $\bar{\delta} \approx 1,9$, and $2/\Psi
           \approx 4.6$ (slopes marked by dashed lines). These values
           have a large uncertainty, as changes in the value of $t_0$
           severely affect them. Estimating these exponents with
           larger precision is computationally very demanding.}
\end{figure}

\begin{figure}[h!]
         \centering
    \includegraphics[width=0.48\textwidth]{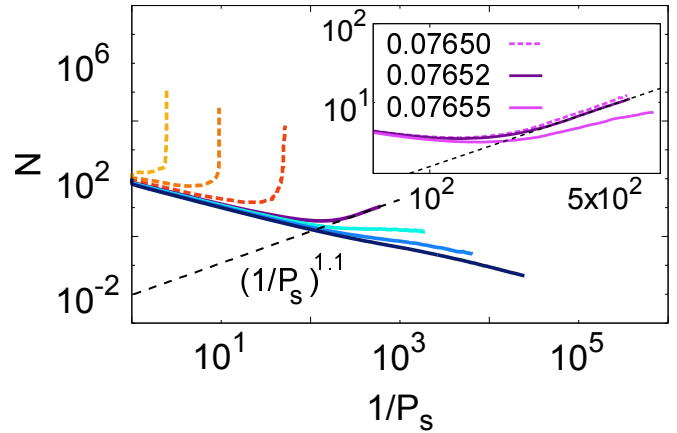}
    \caption{\textbf{(Color online) Double logarithmic plot of $N(t)$
        as a function of $1/P_s(t)$ for spreading experiments at
        criticality in the disordered model \cite{Vojta-Review}.} Our
      best estimate for the slope of at the critical point (separatrix
      of the curves, see main panel) is compatible with the value
      reported in the literature $N(t) \sim
      P_s(t)^{-\bar{\theta}/\bar{\delta}\approx -1.08}$ corresponding
      to the universality class of the directed percolation with
      quenched disorder. The inset shows a zoom around the critical
      point.  Lattice size $N=1024^2$; averages up to $5\times10^7$
      realizations, and same parameters as in Figure 7.}
\label{fig:8} 
\end{figure}

\section{Conclusions and Discussion}
\label{conclusions}

In contrast with the pure model, in the disordered case we have found
a Griffiths phase and a second-order phase transition with an
activated type of scaling. Therefore, in this non-equilibrium system
with one absorbing state the situation remains much as in equilibrium
situations: disorder annihilates discontinuous transitions and induces
criticality.

Results are rather similar to those reported for the standard contact
process with quenched disorder.  Indeed, results are fully compatible
(up to numerical precision) with the standard strong-disorder fixed
point of the universality class of the directed percolation with
quenched disorder \cite{Hoo1,Vojta-Review,Vojta2D}.  We believe that
our results are robust upon considering other types of (weaker)
disorders \cite{Hoyos2008}. Thus, two different models with
significantly different pure versions --i.e. one with a first-order
and one with a second-order transition-- become very similar once
quenched disorder is introduced. Both exhibit Griffiths phases and
activated scaling at the transition point.

From a more general perspective, deciding whether novel universal
behavior emerges in disorder-induced criticality is still an open
problem in statistical mechanics. For illustration, let us point out
that recent work suggests that disorder-induced second-order phase
transitions in an Ising-like system with up-down symmetry does {\it
  not} coincide with Ising transition \cite{Bellafard}. Similarly, in
Ref. \cite{Hoyos2012} a novel type of critical behavior is found for
disorder-induced criticality.  In the case studied here, the
disorder-induced criticality does not seem to lead to novel behavior
(up to numerical precision); indeed, all evidences suggest that it
behavior coincides with the universality class of the directed
percolation with quenched disorder.

After a careful inspection of the literature in search of
discontinuous transitions in disordered non-equilibrium
low-dimensional systems, we found a very recent work in which the
authors study the popular (two-dimensional) Ziff-Gulari-Barshad (ZGB)
model for catalytic oxidation of carbon monoxide \cite{ZGB} in the
presence of catalytic impurities (a fraction of inert sites)
\cite{Buendia}.  The pure ZGB model is known to exhibit, among many
other relevant features, a discontinuous transition into an absorbing
state.  However, after introducing quenched disorder, no matter how
small its proportion, the discontinuous transition is replaced by a
continuous one \cite{Buendia}, similarly to our findings here.

In conclusion, we conjecture that first-order phase transitions cannot
appear in low-dimensional disordered systems with an absorbing
state. In other words, the Imry-Ma-Aizenman-Wehr-Berker argument for
equilibrium systems can be extended to non-equilibrium situations
including absorbing states. The underlying reason for this is that,
even if the absorbing phase is fluctuation-less and hence is free from
the destabilizing effects the Imry-Ma argument relies on, the other
phase is active and subject to fluctuation effects. Therefore,
intrinsic fluctuations destabilize it as predicted by the
Imry-Ma-Aizenman-Wehr-Berker argument, precluding phase coexistence.

Remarkably, in the case studied by Barghathi and Vojta, in which the
Imry-Ma argument is violated in favor of a second-order phase
transition, the two broken-symmetry states are absorbing ones: once
the symmetry is broken in any of the two possible ways, the system
becomes completely frozen, i.e.,  free from fluctuation effects, and,
consequently, the Imry-Ma argument breaks down.  Thus, the only
possibility to have first-order phase transitions in low-dimensional
disordered systems would be to have (in its pure version counterpart)
a discontinuous phase transition between two different
fluctuation-less states, and we are not aware of any such
transition. Therefore, we conclude that quenched disorder forbids
discontinuous phase transitions in low-dimensional non-equilibrium
systems with absorbing states.

\begin{acknowledgments}
  We acknowledge support from the J. de Andaluc\'{i}a project of
  Excellence P09-FQM-4682, the Spanish MEC project FIS2009--08451,
  schollarship FPU2012/05750, and NSF under grant OCE-1046001. We
  thank P. Moretti for a careful reading of the manuscript.
\end{acknowledgments}

\appendix

\section{Simple mean field approach} 

In order to obtain a simple mean-field approach to the present
problem, expected to hold in the high-dimensional limit, we consider a
fully connected lattice in which each node is nearest neighbor to any
other one. $n(t)$ is the total number of particles at a given time and
$\rho(t)=n(t)/N$ the corresponding density.  Allowed changes at any
time-step are of the magnitude $ \pm 1/N$.  We can thus write
the overall transition rates as $ W^-(\rho \to \rho-1/N) =\mu \rho $
%\nonumber \\
and
$W^+(\rho \to \rho+1/N)  = \lambda \rho^2(1-\rho)$
%\nonumber \\
%W^0(\rho \to \rho,t)& =1-[W^-(\rho,t)+W^+(\rho,t)] \nonumber
%\end{align}
for creation and annihilation processes, respectively.  Expanding the
associated master equation in power series, and keeping only the first
two leading terms we obtain the Fokker-Planck equation:
\begin{align}
\partial_t P(\rho,t) 
%& =-\partial_\rho \left \{ [W^+(\rho,t)-W^-(\rho,t)]P(\rho,t) \right \} + \nonumber \\
%& +\displaystyle\frac{1}{2N} \partial_\rho^2 \left \{
 % [W^+(\rho,t)+W^-(\rho,t)]P(\rho,t) \right \} \nonumber  \\
& =-\partial_\rho[(-\lambda\rho^3+\lambda\rho^2- \mu \rho)P] \nonumber + \\
  & +\displaystyle\frac{1}{2N}\partial_\rho^2[(-\lambda\rho^3+\lambda\rho^2+\mu \rho) P] \nonumber,
\end{align}
equivalent to the (It\v o) Langevin equation \cite{Gardiner}:
\begin{align}
  \partial_t\rho =-\mu \rho +\lambda\rho^2 -\lambda\rho^3
  +\sqrt{\displaystyle\frac{\mu \rho-\lambda\rho^3+\lambda\rho^2}{N}}\xi(t). \nonumber
\end{align}
where $\xi(t)$ is a Gaussian white noise.  In the $N\rightarrow
\infty$ (mean-field) limit one recovers the rate equation 
[Eq.(\ref{rate})] with its associated discontinuous transitions. Using
the noise term it is possible to derive (using the theory of mean
first passage times \cite{Gardiner,Meerson,Lubensky}) the scaling of
the escape time as a function of the system size.

In general, one-component reaction diffusion systems with $l$-particle
creation and $k$-particle annihilation \cite{Kamenev}:
\begin{equation}\label{general-model}
  k A {\rightarrow} (k-n) A  \,,\quad \,\,
  l A {\rightarrow} (l+m) A  \,,\quad 
\end{equation}
always exhibit a first-order transition if $l > k$ if the particles
are fermionic. Instead, if the  hard-core constraint is excluded, an additional
reaction $i A {\rightarrow} j A$ with $i>j$ and $i>l$ is needed to
stabilize the system.

\def\url#1{}

\end{document}